\documentclass[referee]{raa}            

\usepackage{graphicx,times}             
\usepackage{natbib}
\usepackage{amssymb,amsmath}
\usepackage{amssymb}
\bibpunct{(}{)}{;}{a}{}{,}

\usepackage[a4paper=true,pagebackref=true]{hyperref}
\hypersetup{colorlinks = true, linkcolor = green, anchorcolor = red, citecolor = blue, filecolor = red, pagecolor = red, urlcolor = red}

\begin{document}

   \title{The Likely Counterpart to $\gamma$-Ray Excess from The Northwest Region of Arp 220
}

   \volnopage{Vol.0 (20xx) No.0, 000--000}     
   \setcounter{page}{1}          
   \author{Yunchuan Xiang
      \inst{1}
   \and   Junhao Deng
      \inst{1}
   \and Zejun Jiang
      \inst{1}
   }

   \institute{Department of Astronomy, Yunnan University, and Key Laboratory of Astroparticle Physics of Yunnan Province, Kunming, 650091, China, xiang{\_}yunchuan@yeah.net, zjjiang@ynu.edu.cn\\
   {\small Received~~19-Feb-2021; accepted~~27-Feb-2021}}

\abstract{
The unknown $\gamma$-ray excess in the northwest region of Arp 220 
was revisited by analyzing $\sim$11.8 years of the \textit{Fermi} Large Area Telescope  (\textit{Fermi}-LAT) data in this study. We found that its photon flux was  approximately three times higher than that of the previous study in the 0.2-100 GeV band, and the  corresponding significance level$\sim8.15\sigma$ was approximately four times higher than before.
The light curves of 15 and 45 time bins from the whole time all showed two active periods,  and the variability of the second period was more significant than that of the first period.
The spectral indices from the two active periods were not statistically different and were close to the range of $\gamma$-ray flat-spectrum radio quasars observed by \textit{Fermi}-LAT.
Because the position of CRATES J153246+234400 was consistent with the best-fit position of our analysis, we suggest that CRATES J153246+234400 is more likely a $\gamma$-ray counterpart for the variational region. 
For Arp 220, there was no significant variability in the $\gamma$-ray emission.
}

   \authorrunning{Y.C. Xiang, J.H. Deng \& Z.J. Jiang}            
   \titlerunning{ The GeV $\gamma$-ray Emission of The Northwest Region from Arp 220}  

   \maketitle

%
%
\section{Introduction}
\label{sect:intro}

Arp 220, as the nearest ultraluminous infrared galaxy,  has a total infrared luminosity of $(1-2)\times10^{12} L_{\bigodot}$ \citep{Sanders2003,Gao2004,Rangwala2011}. 
 \citet{Peng2016} first found the GeV emission of this galaxy using 7.5 years of \textit{Fermi} Large Area Telescope (\textit{Fermi}-LAT) observations.  
They analyzed the light curve (LC) of 7.5 years of the $\gamma$-ray emission of Arp 220, and no significant variability of the LC was found in their results. The luminosity of $L_{\rm 0.1-100 \ GeV}$=(1.78$\pm$0.30)$\times 10^{42}\  erg\  s^{-1}$ for Arp 220 is in accordance the quasilinear scaling relation between the total infrared luminosity and $\gamma$-ray luminosity for starburst galaxies and star-forming galaxies \citep{Thompson2007,Ackermann2012,Peng2016}. 
It is worth noting that \citet{Peng2016} found the residual $\gamma$-ray emission in the northwest region of Arp 220. After analyzing the source of the radiation separately, they suggested that the residual $\gamma$-ray emission was likely to originate the single or collective contribution of four active galactic nuclei (AGNs) candidates.

In fact, \citet{Tang2014} reported the $\gamma$-ray excess of the northwest region of Arp 220 and suggested that the contribution of the $\gamma$-ray excess was more likely to come from CRATES J153246+234400, for a small angular separation of 0$^{\circ}$.16 away from CRATES J153246+234400. 
Moreover, they found significant variability between the highest and lowest annual fluxes from an annual LC of the $\gamma$-ray emission.

CRATES J153246+234400, as a flat-spectrum radio quasar (FSRQ),  is also known as ICRF J153246.3+234405 with an unknown redshift.  
CRATES J153246+234400 is from the catalog of radio sources discovered at the frequency of 4.8 GHz; its  
4.8-GHz flux density is 109 mJy with spectral index $\sim$0.213, and its morphology class is flagged as a point source \citep{Healey2007}. 
FSRQs, a subclass of blazar, are generally believed to have Doppler-boosted relativistic jets that point toward the line of sight from observers \citep{Blandford1978,Urry1995}.
FSRQs are radio-loud AGNs, and 
their jets can produce high-energy emissions from radio to $\gamma$-ray energies for the Doppler beaming effect \citep{Urry1995}. Unlike BL Lacertae objects (BL LACS), FSRQs have strong and broad  optical lines \citep{Stickel1991}.  In general, the GeV $\gamma$-ray emission of FSRQs is considered to originate from the inverse Compton scattering process \citep{Konigl1981,Band1985,Maraschi1992,Blazejowski2000}.

With the accumulation of photon numbers of Arp 220 from \textit{Fermi}-LAT, we reanalyzed the $\gamma$-ray emission of the Arp 220 region.
Through a preliminary analysis of the global fit of  approximately 12 years of \textit{Fermi}-LAT data, we found that the significant variability from  
the northwest region was higher than that reported by \citet{Tang2014} and \citet{Peng2016}, and the variational result strongly inspired us to explore the unknown origin and characteristics of $\gamma$-ray radiation for the northwest region of Arp 220.

This paper is organized as follows:  The data analysis routines and analysis results are provided in Section 2.  Section 3 presents the discussion and conclusion of this work.

\section{Data preparation}
\label{sect:Data}
CRATES J153246+234400 was considered close to the best-fit position (R.A.=233$^{\circ}$.24, decl.=23$^{\circ}$.80)$\pm$10$^{\\'}$.08, and is within a 1$\sigma$ error circle from the calculation result of \citet{Peng2016}. Moreover, \citet{Tang2014}  believed that the $\gamma$-ray excess likely originated from CRATES J153246+234400. 
Therefore, we selected the position of CRATES J153246+234400 to analyze the residual  $\gamma$-ray radiation from the northwest region of Arp 220.

The time range of the photon events was selected from 2008-08-04 to 2020-05-14 (MET 239557427-MET 611174172) for this analysis, and the energy range was 0.1-500 GeV. 
A $20^{\circ}\times 20^{\circ}$ region of interest (ROI) centered at the position (R.A.=233$^{\circ}$.19, decl.=23$^{\circ}$.73) from SIMBAD\footnote{ http://simbad.u-strasbg.fr/simbad/} for CRATES J153246+234400.
The version {\tt v11r5p3} package\footnote{http://fermi.gsfc.nasa.gov/ssc/data/analysis/software/} of the \textit{Fermi} Science Tools was used to perform this analysis.
The data analysis method from \textit{Fermi} Science Support
 Center\footnote{http://fermi.gsfc.nasa.gov/ssc/data/analysis/scitools/} was followed.
For event class from the Pass 8 data, ``Source'' event class with evclass = 128 and evtype = 3 was also selected. Considering the contamination from the Earth limb, we excluded events with zenith angles $>90^{\circ}$.
For the instrumental response function,  ``P8R3{\_}SOURCE{\_}V2'' was adopted. 
 We divided all data of the northwest region of Arp 220 into 37 logarithmic energy bins with a $0^{\circ}.1\times0^{\circ}.1$ spatial resolution.
The binned maximum likelihood method was used to fit the data with two diffuse backgrounds, including the isotropic extragalactic emission ({\tt iso{\_}P8R3{\_}SOURCE{\_}V2{\_}v1.txt}) 
and the galactic diffuse emission ({\tt gll{\_}iem{\_}v07.fits})\footnote{http://fermi.gsfc.nasa.gov/ssc/data/access/lat/BackgroundModels.html}. 
Other known sources from the \textit{Fermi} Large Telescope Fourth Source Catalog\footnote{https://fermi.gsfc.nasa.gov/ssc/data/access/lat/10yr{\_}catalog/} (4FGL) within the ROI were included in the model file, which has 316 objects
within 30$^{\circ}$, centered at the position of CRATES J153246+234400. All spectral parameters and normalizations from these sources within the 5$^{\circ}$ region centered at the position of CRATES J153246+234400 were set as free, and the normalizations of two diffuse backgrounds were also set as free.

\subsection{\rm Certification of the $\gamma$-ray source} \label{sec:data-reduction}
A point source with a power-law spectral model at the position of CRATES J153246+234400 was added to the model file to analyze its $\gamma$-ray emission.
To derive the best-fit position of the $\gamma$-ray source from CRATES J153246+234400,  the command {\tt gtfindsrc} was used, and the best-fit position was R.A. = 233$^{\circ}$.19, decl. = 23$^{\circ}$.72 with 1$\sigma$ error radius of 0$^{\circ}$.04.
The separation between the position of CRATES J153246+234400 and the best-fit position of the $\gamma$-ray source was only 0$^{\circ}$.02, indicating that CRATES J153246+234400 was within the error circle with a 68\% confidence level. 
In the model file, the position of CRATES J153246+234400 was replaced with the best-fit position of the $\gamma$-ray emission for all the subsequent analyses. 
Using the binned maximum likelihood method, we found that the photon flux of the global fit of CRATES J153246+234400 was  $(1.63\pm 0.18) \times 10^{-8} \rm  cm^{-2}\ s^{-1}$ with the  spectral index of $\sim$ 3.11$\pm$0.14, and test statistic (TS) value $\sim$ of 184.95. 

\begin{figure*}[!h]
  \begin{minipage}[t]{0.495\linewidth}
  \centering
   \includegraphics[width=70mm]{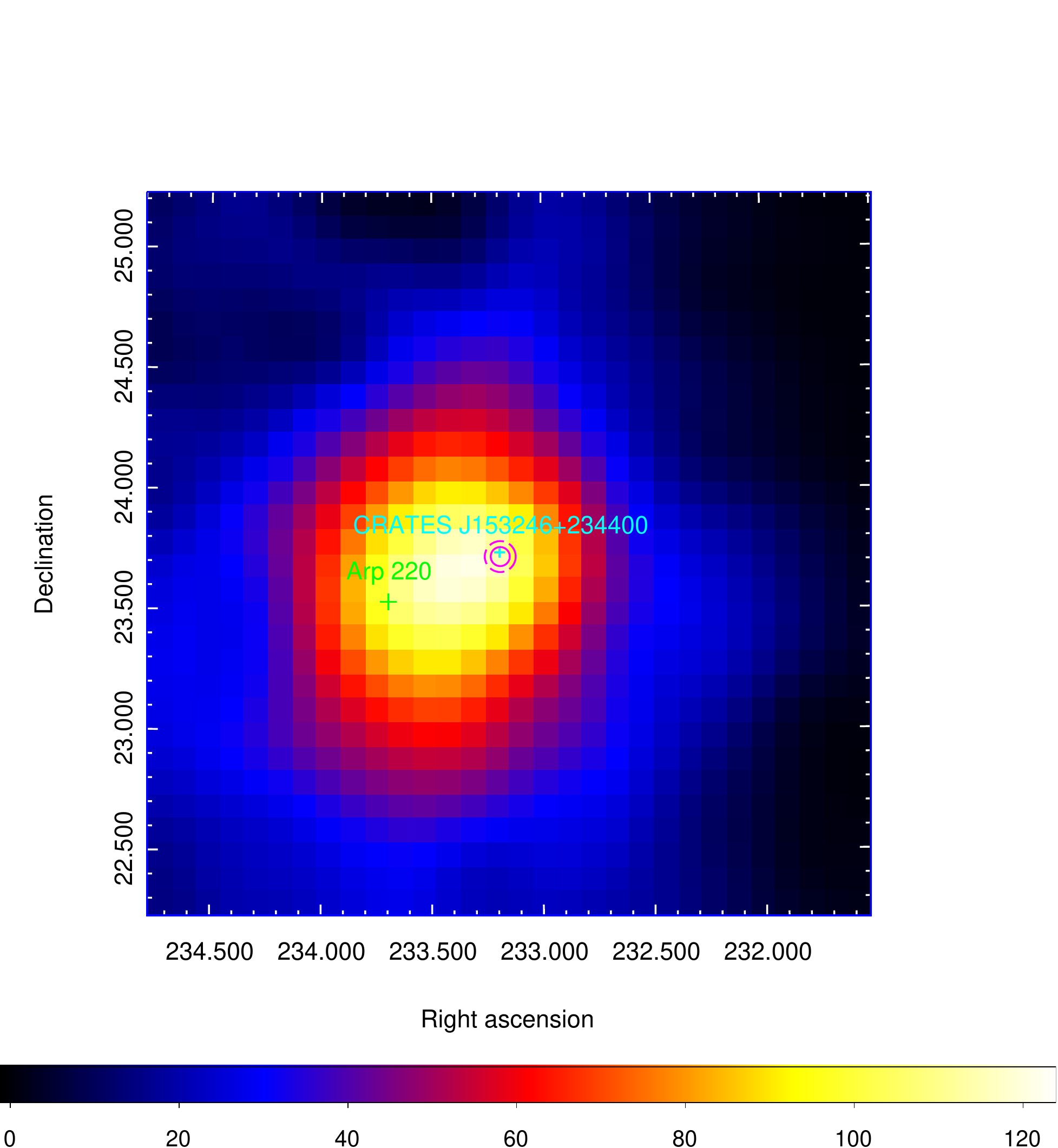}
   \centerline{(\emph{a})}
  \end{minipage}%
  \begin{minipage}[t]{0.495\textwidth}
 \centering
  \includegraphics[width=70mm]{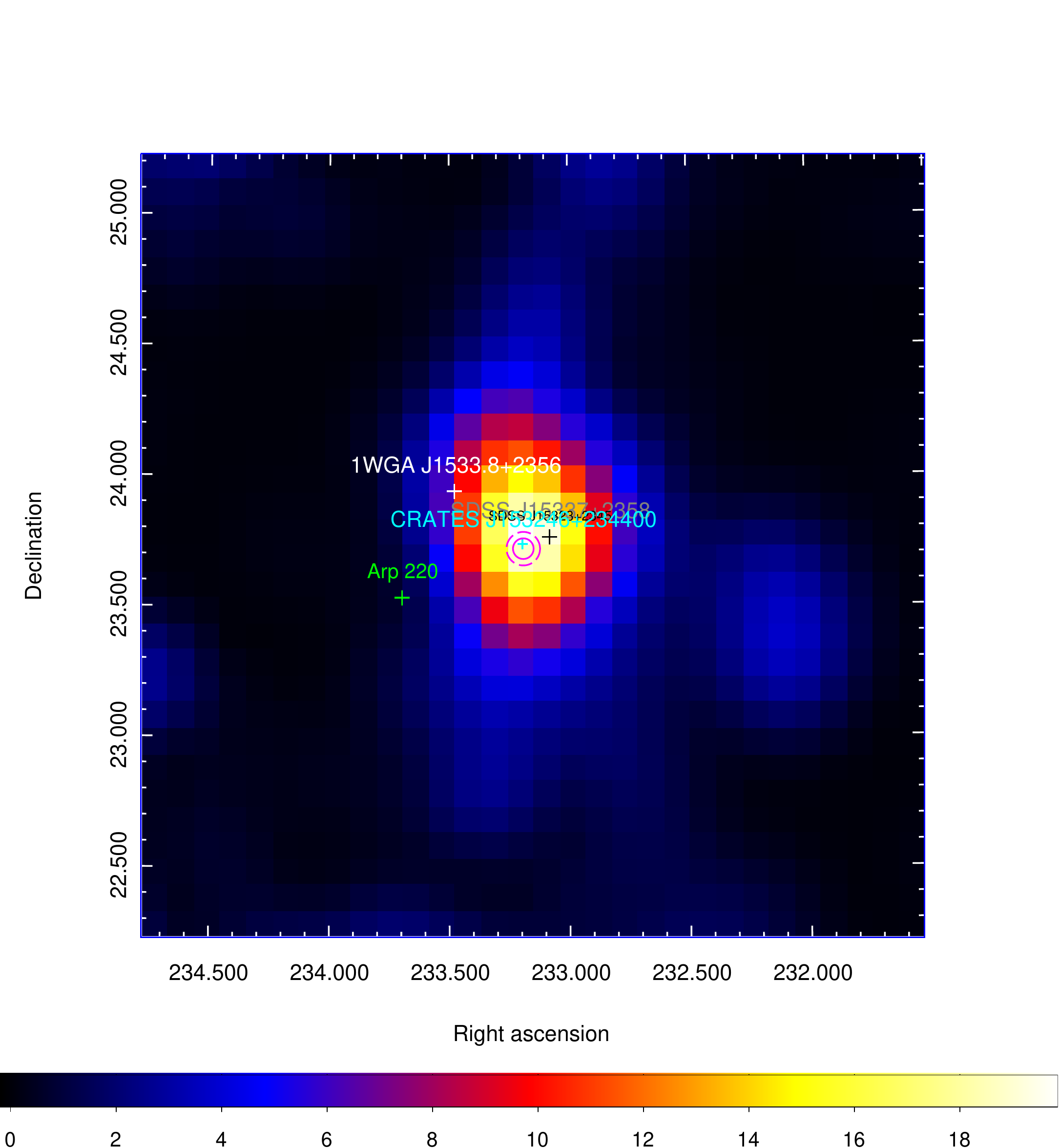}
    \centerline{(\emph{b})}
 \end{minipage}%
 \vfill
  \begin{minipage}[t]{1\linewidth}
  \centering
  \includegraphics[width=70mm]{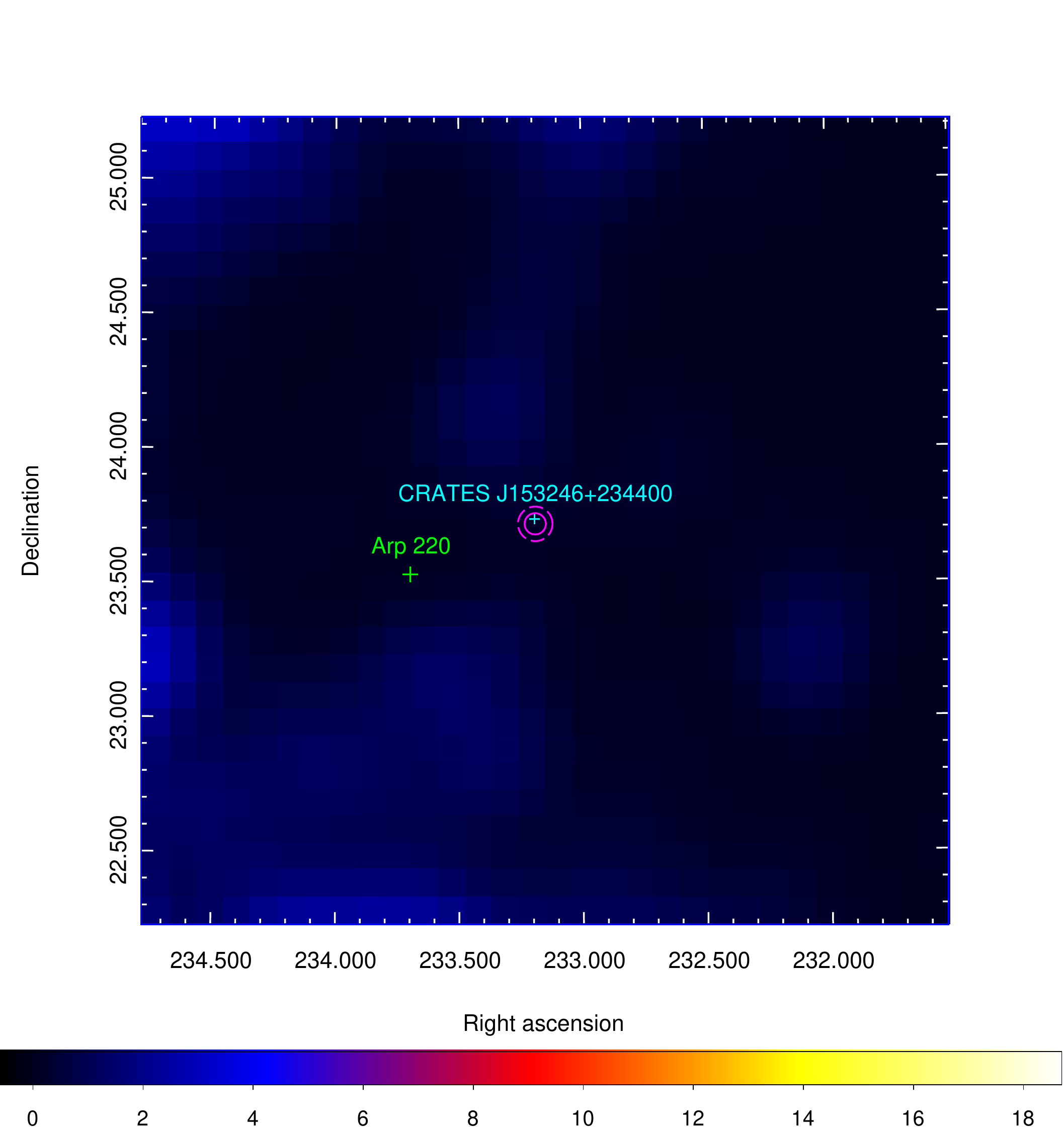}
    \centerline{(\emph{c})}   
 \end{minipage}%
\caption{TS maps for a $3^{\circ}\times3^{\circ}$ region centered at the best-fit position of CRATES J153246+234400. These TS maps were smoothed with the Gaussian kernel of $\sigma =0^{\circ}.3$ and created using a grid of $0^{\circ}.1$. The cyan crosses represent the position of CRATES J153246+234400 from SIMBAD, and the magenta circles show the 68\% (inner) and 95\% (outer) error regions of the best-fit position of CRATES J153246+234400 for these TS maps. The green crosses represent the position of Arp 220 from \citep[4FGL;][]{Abdollahi2020}.
 Panel (\emph{a}) shows the TS map where they all existed. Panel (\emph{b}) shows the residual TS map deducting Arp 220, where the black and gray crosses represent the positions of SDSS J15323+2345 and SDSS J15337+2358, respectively, and the white cross represents the position of 1WGA J1533.8+2356. Panel (\emph{c}) shows the residual TS map, where the two sources were excluded.
   }
    \label{TSmap}
\end{figure*}

The TS map for a $3^{\circ}\times3^{\circ}$ region centered at the best-fit position of CRATES J153246+234400 was first generated running {\tt gttsmap} for the entire time in the 0.2-500 GeV energy band.
To compare the variability of the GeV emission from the  region of CRATES J153246+234400 between the results of  \citet{Peng2016} and our study,  we first retained the model of 4FGL J1534.7+2331 associated with Arp 220 in the model file. A more significant source with the TS value=89.02 was found toward CRATES J153246+234400 above 200 MeV compared with the result of \citet{Peng2016}, as shown in panel (\emph{a}) in Figure \ref{TSmap}, 
and the significance level of $\gamma$-ray emission from the northwest region of Arp 220 was higher $\sim$4 times than that of \citet{Peng2016}.
On this basis, we subtracted the $\gamma$-ray emission from 4FGL J1534.7+2331, the TS map was again generated, and a significant $\gamma$-ray excess still existed in the region of CRATES J153246+234400, as shown in panel (\emph{b}) of Figure \ref{TSmap}. 
After subtracting all sources including CRATES J153246+234400, as shown in panel (\emph{c}) of Figure \ref{TSmap},  we found that there was no significant $\gamma$-ray emission excess from the TS map, suggesting that the $\gamma$-ray excess was likely to come from CRATES J153246+234400 in this analysis, which is consistent with the results of \citet{Tang2014}.

\citet{Peng2016} performed a likelihood analysis for this region in their study;  their calculated best-fit position was (233$^{\circ}$.24, 23$^{\circ}$.80)$\pm$0$^{\circ}$.17, while our obtained best-fit position was (233$^{\circ}$.19, 23$^{\circ}$.72)$\pm$0$^{\circ}$.04. 
\citet{Peng2016} checked likely candidates from three AGN catalogs for the region in the vicinity of $r_{95}$ from the best-fit position from their analysis.
They provided four likely candidates containing CRATES J153246+234400, and the other three objects were SDSS J15337+2358, SDSS J15323+2345, and 1WGA J1533.8+2356.   
We found that these last three sources were all located outside of the 95\% error circle of the best-fit position of our analysis, as shown in  panel (b) of Figure \ref{TSmap}.
Therefore, we suggest that the $\gamma$-ray emission from this region is more likely to originate from CRATES J153246+234400.

\subsection{Light Curve}

For the result of the global fit in 0.2-100 GeV from the region of CRATES J153246+234400 from \citet{Peng2016}, the photon flux was $(1.45\pm0.52)\times 10^{-9}$ ph cm$^{-2}$ s$^{-1}$ with  the spectral index of $\sim$ 2.45$\pm$0.19, and the TS value was 22.
With the accumulation of photon numbers from \textit{Fermi}-LAT, the TS value of the region of CRATES J153246+234400 increased to 80.50,  with 8.27$\sigma$, from our analysis, and its photon flux was $(2.68\pm 0.60)\times10^{-9}$ ph cm$^{-2}$ s$^{-1}$, with the spectral index of $\sim$2.63$\pm$0.14 in the 0.2-100 GeV energy band.
Compared with the results of \citet{Peng2016},
we found that our photon flux was approximately two times larger, and the TS value was approximately four times larger, implying that CRATES J153246+234400 was a likely $\gamma$-ray variability source.
To confirm the likely variability from the $\gamma$-ray emission of CRATES J153246+234400, an LC of CRATES J153246+234400 was generated with energy above 200 MeV to avoid a large point spread function (PSF) in the lower energy band.
Here, we generated the LC with 15 linearly equal time bins for the entire time. The maximum-likelihood analysis method was used separately to fit each time bin. 
 For a photon flux of the TS value $<$ 4, a 95\% flux upper limit was calculated.

\begin{figure*}[!h]
  \begin{minipage}[t]{0.495\linewidth}
  \centering
   \includegraphics[width=70mm]{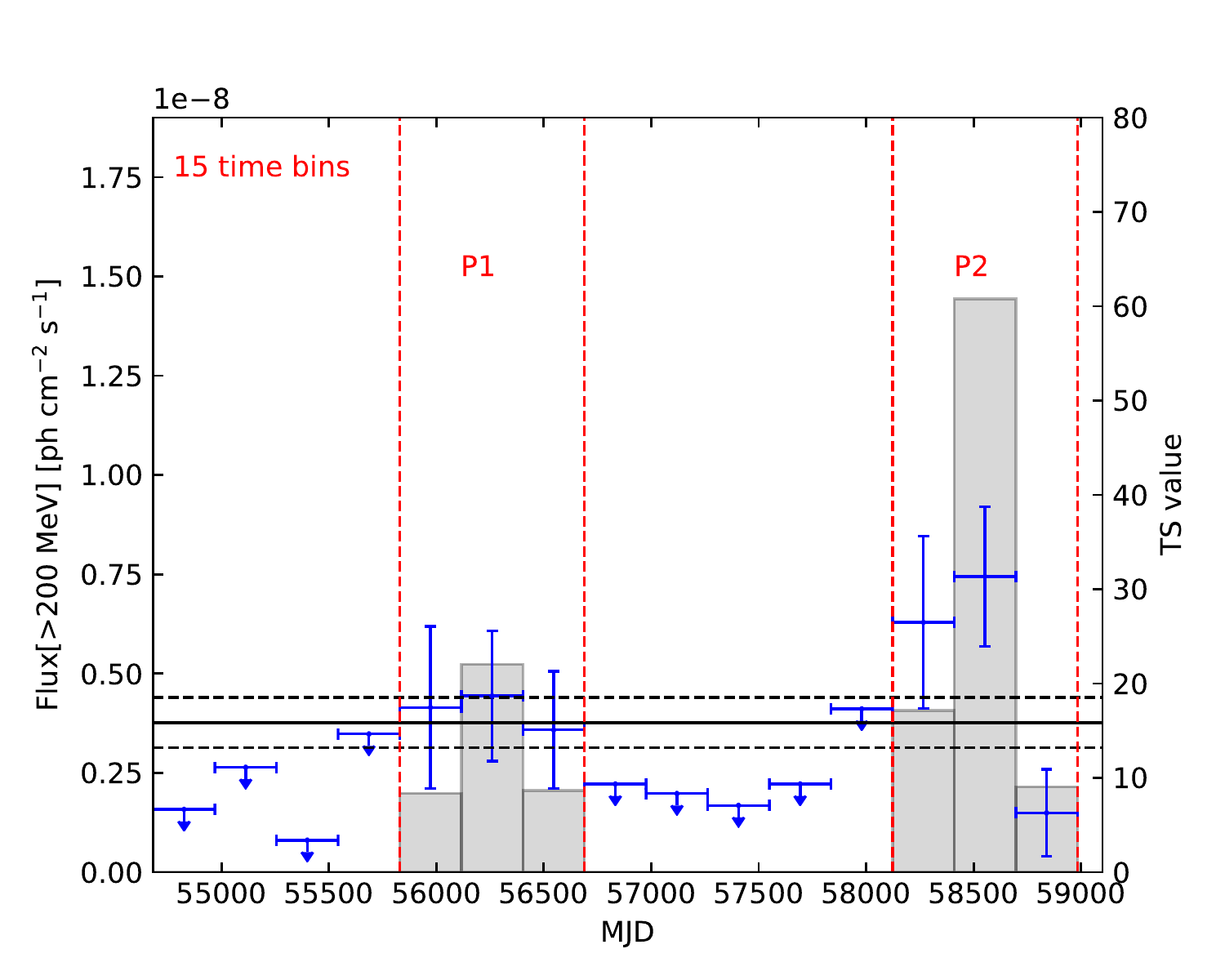}
   \centerline{(\emph{a})}
  \end{minipage}%
  \begin{minipage}[t]{0.495\textwidth}
 \centering
  \includegraphics[width=70mm]{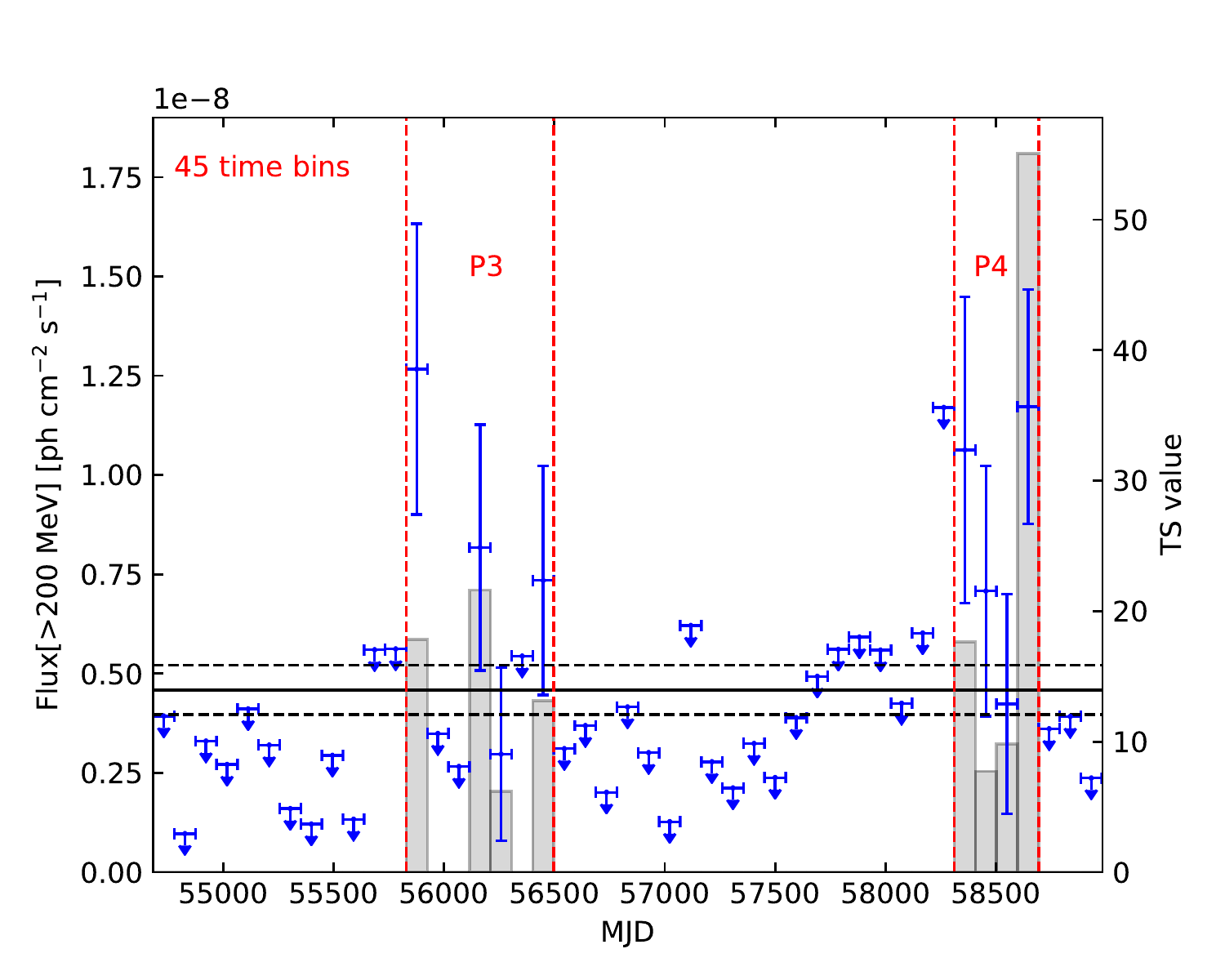}
    \centerline{(\emph{b})}
 \end{minipage}%
 \vfill
\caption{LCs of 15 and 45 time bins of $\gamma$-ray emission toward CRATES J153246+234400 from 0.2 GeV to 500 GeV. The gray shaded regions show the TS value of each time bin.  The red dashed lines divide the two active periods for the two LCs, which the two active periods are marked as P1, P2, P3, P4. The black solids and the dashed lines represent the best-fit results and 1$\sigma$ uncertainties of a constant flux model for the time bin of the TS value $>$ 1, respectively.
}
    \label{LC}
\end{figure*}

The $\gamma$-ray photon flux and index in the 0.2-500 GeV band for each time bin with a TS value$>$1 were all fitted using a constant flux model \citep[e.g.,][]{Zhang2016,Peng2019}, as shown in panel (\emph{a}) in Figure \ref{LC}; the fitting results of photon flux and the reduced $\chi^{2}$ were 
$(3.76\pm 0.63) \times 10^{-9} \rm  cm^{-2}\ s^{-1}$ and 1.76, respectively. Then, we found that the photon data  were mainly concentrated in two active periods of P1 and P2 , represented by the red dashed lines in panel (\emph{a}) of Figure \ref{LC}.
For the LC with 15 time bins, we found that the values of the photon flux and TS from the middle bins of P1 and P2 were higher than those of the other bins. Moreover, the value of the photon flux of the middle bin of P2 was $\sim$2 times higher than that of the 1$\sigma$ uncertainty of the constant flux model, which also implied the likely variability from the LC.

Subsequently, a finer LC of 45 time bins with photon energy $>$ 200 MeV was generated to further check whether there still existed a significant variability of the photon flux over the entire time, as shown in panel (\emph{b}) of Figure \ref{LC}. Using the fit method of a constant flux model, as was done in the LC with the 15 time bins, the best-fit results of the photon flux and the reduced $\chi^{2}$ of the LC with 45 time bins were
 $(4.95\pm 0.62)\times 10^{-9}\rm  cm^{-2}\ s^{-1}$ and 4.84, respectively.
The related best-fit results of the constant flux model for the cases of 15 and 45 time bins are listed in Table \ref{Table1}. Meanwhile, 
periods of P1 and P2 were compressed into periods P3 and P4, as shown in panel (\emph{b}) of Figure \ref{LC}.

We found that the TS values from the two active periods of P3 and P4 were higher than those of the other time bins. 
Moreover, the six values of photon flux with 1$\sigma$ statistical uncertainty from P3 and P4 were significantly higher than the 1$\sigma$ uncertainty of the constant flux model in the LC with 45-time bins, and the photon flux of the fourth bin of P4 had a higher photon flux of $(1.17\pm 0.30) \times 10^{-8} cm^{-2}\ s^{-1}$ and the TS value of 55.08 than the other time bins, which further implied that the variability from the photon flux of CRATES J153246+234400 could not be overlooked.

To further evaluate the variability of the LCs with 15 and 45 time bins, the variability index ($\rm TS_{\rm var}$) defined by \citet{Nolan2012} was calculated in our analysis. 
For the LC of 15 time bins,
 the value of $\rm TS_{\rm var}\geq$29.14 was used to distinguish variable sources at a 99\% confidence level. The value of $\rm TS_{\rm var}=60.95$ with 5.37$\sigma$ for the LC of 15 time bins suggests the variability from the photon flux of CRATES J153246+234400.
Similarly, for the LC of 45 time bins, the value of $\rm TS_{\rm var}\geq$68.71 was used to distinguish variable sources at a 99\% confidence level, and $\rm TS_{\rm var}=122.04$ with 5.94$\sigma$ also indicates that the variability of the photon flux of CRATES J153246+234400 does exist.

\begin{figure*}[!h]
  \begin{minipage}[t]{0.495\linewidth}
  \centering
   \includegraphics[width=70mm]{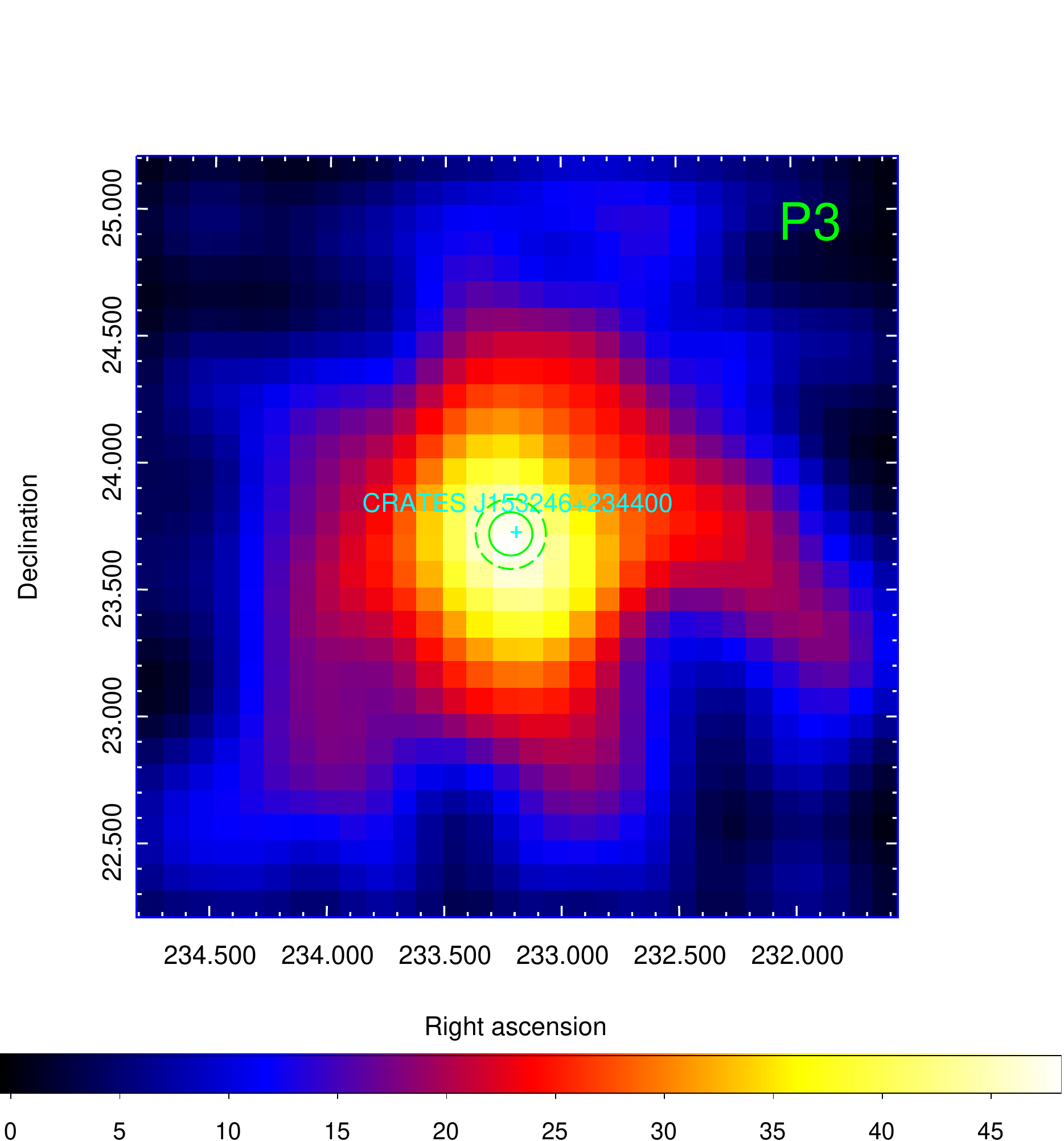}
   \centerline{(\emph{a})}
  \end{minipage}%
  \begin{minipage}[t]{0.495\textwidth}
 \centering
 
  \includegraphics[width=70mm]{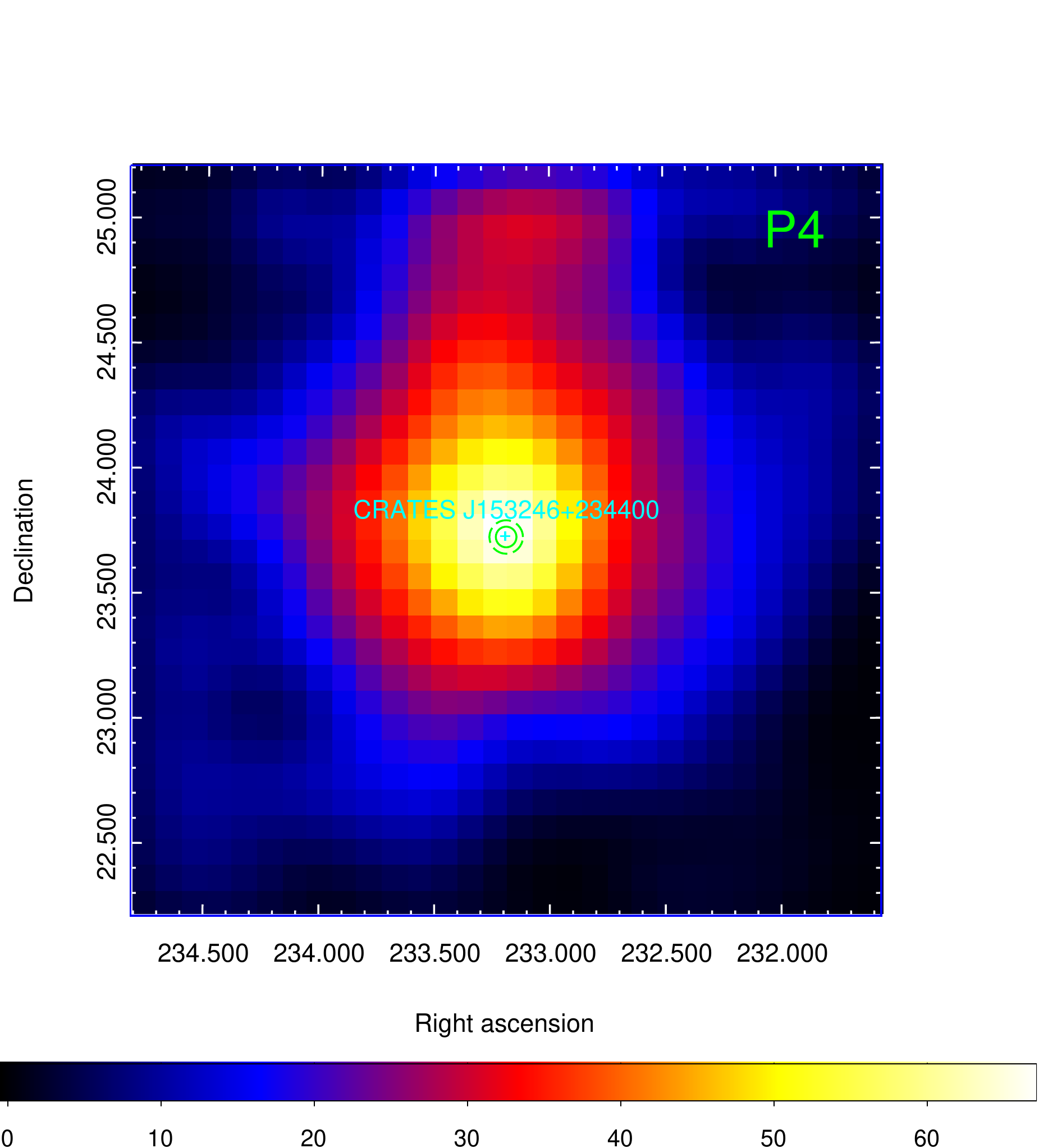}
    \centerline{(\emph{b})}
 \end{minipage}%
\caption{
Two panels show $3^{\circ} \times 3^{\circ}$ TS maps with a grid of $0^{\circ}.1$ smoothed with the Gaussian kernel of $\sigma =0^{\circ}.3$ centered at the best-fit position of CRATES J153246+234400. 
Panels (a) and (b) show the TS maps at P3 and P4, respectively.
 The cyan crosses of the two panels represent the position of CRATES J153246+234400, and the cyan circles show the 68\% (inner) and 95\% (outer) error regions of the best-fit positions of CRATES J153246+234400 from P3 and P4, respectively.}
    \label{Fig3}
\end{figure*}

To further confirm whether the $\gamma$-ray excesses of P3 and P4 originate from CRATES J153246+234400,
we obtained uncertainties of the 68\% and 95\% of the corresponding best-fit position of CRATES J153246+234400 from P3 and P4, respectively.  TS maps of $3^{\circ}\times 3^{\circ}$ from P3 and P4 centered at the best-fit position of CRATES J153246+234400 were generated. 
As shown in Figure \ref{Fig3}, we found that the position of CRATES J153246+234400 was always within the 68\% error circles of the best-fit positions of CRATES J153246+234400 for the two different active periods of P3 and P4. The best-fit positions and 68\% error circles, marked as $r_{68}$, of CRATES J153246+234400 from P3 and P4 are provided in Table \ref{Table2}.
These results strongly confirm that the $\gamma$-ray excesses from P3 and P4 probably came from CRATES J153246+234400.

 \begin{table}[!h]
 \scriptsize
\begin{center}
\caption[]{fitting parameters with a constant flux model for the two different LCs}
\begin{tabular}{clclclclclclclc}
  \hline\noalign{\smallskip}
    \hline\noalign{\smallskip}
      Light Curve    & Fitting Flux               & $\chi_{\rm red}^{2}$ & $\rm TS_{\rm var}$ & Significance Level\\
                     & $10^{-9}$ cm$^{-2}$ s$^{-1}$ &                      &               &                     &                &         \\
  \hline\noalign{\smallskip}
  15 time bins      & 3.76$\pm$0.63                  & 1.76    & 60.95 & 5.37$\sigma$ \\
  45 time bins      & 4.59$\pm$0.62                   & 4.84     & 122.04 & 5.94$\sigma$  \\
   \hline\noalign{\smallskip}
\end{tabular}

\textbf{Note}: the $\chi_{\rm red}^{2}$ represents the reduced $\chi^{2}$ of fitting photon flux calculated by a $\chi^{2}$ minimization procedure of a constant flux model.
 \label{Table1}
\end{center}  
\end{table}

 \begin{table}[!h]
 \scriptsize
\begin{center}
\caption[]{The best-fitting parameters with the PL model for the two \textbf{active} periods}
\begin{tabular}{clclclclclclclc}
  \hline\noalign{\smallskip}
    \hline\noalign{\smallskip}
      Active Period & Time Range    & $ F_{100}$   &  $\Gamma$     &  TS Value & Significance Level & The Best-fit Position & $r_{68}$\\
                    & MJD   &$10^{-8}$ cm$^{-2}$ s$^{-1}$  &  &  &  &  degree &   degree   \\
  \hline\noalign{\smallskip}
  P3   & 55828.66-56497.16  & $1.59\pm0.50$        &  $2.56\pm0.17$         &  57.14  &   6.79$\sigma$& (233.22, 23.73) & 0.09 \\
  P4  & 58311.66-58693.66   &$2.17\pm0.76$        &  $2.49\pm0.17$         &  87.17 &   8.65$\sigma$ & (233.19,  23.73) & 0.04 \\
   \hline\noalign{\smallskip}
\end{tabular}
 \label{Table2}
\end{center}
\end{table}

\subsection{Spectral Energy Distribution}
\label{sect:Spectral Energy Distribution}

\begin{figure}[!h]
\centering
 \includegraphics[width=\textwidth, angle=0, scale=0.6]{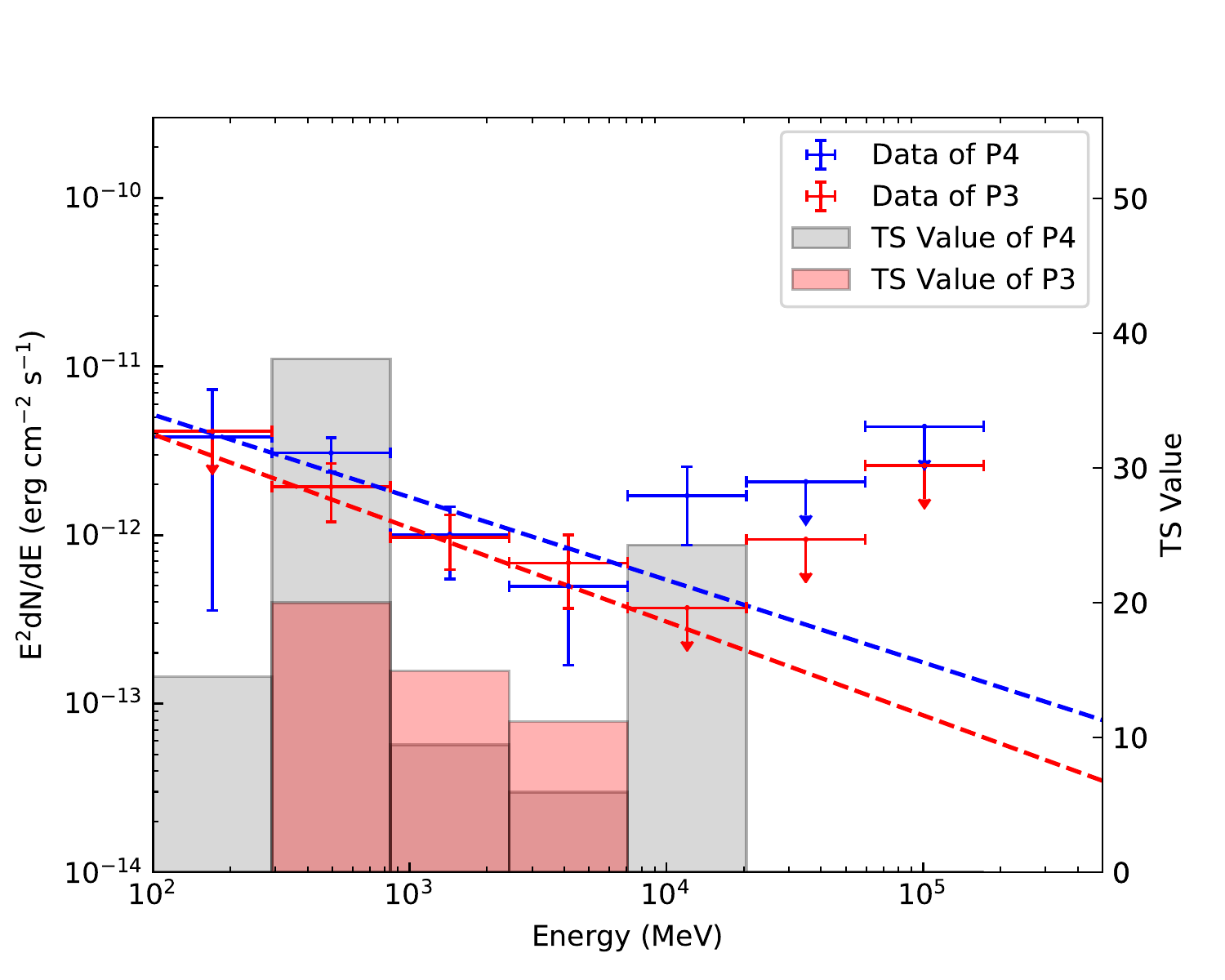}
 \caption{The blue and red data points are from the P4 and P3 periods, respectively. The gray and red shaded regions show the corresponding TS values from the P4 and P3 periods, respectively. The blue and red dashed lines represent the best-fit power-law spectra from the two periods, respectively. The data points with a TS value $<$ 4 were calculated as the upper limit for the two different periods.
 }
 \label{SED}
\end{figure}

The spectral energy distribution (SED) from 100 MeV to 500 GeV was  derived for P3 and P4 using the power-law (PL) model. The entire energy range was divided into eight equally spaced log10(E) energy bins, as shown in Figure \ref{SED}. The data points with a TS value $<$ 4 were calculated as the upper limit at the 95\% confidence level for P3 and P4.
The results of the global fit from P3 and P4 showed that there was no statistical difference in photon fluxes and spectral indices, as shown in Table \ref{Table2}. Furthermore, the significance level of P4 was higher than that of P3.

\subsection{Exploring the GeV flux variability from Arp 220}

With the accumulation of the photon number from \textit{Fermi}-LAT, the position of 4FGL J1534.7+2331 in 4FGL associated with Arp 220 was  selected to reanalyze the related properties of the $\gamma$-ray emission of Arp 220.
Here, the $\gamma$-ray emission of Arp 220 was analyzed using the binned likelihood analysis method with the same time range as the above analysis of CRATES J153246+234400.
The photon flux was $(1.18\pm 0.53)\times 10^{-9}$ ph cm$^{-2}$ s$^{-1}$ with the spectral index $\sim$ 2.29$\pm$0.21 and the TS value $\sim$ 38.43 in 0.2 $\sim$ 100 GeV band, with no statistical difference from those of \citet{Peng2016}.
Moreover, the luminosity of Arp 220 in the 0.1$\sim$100 GeV band was calculated to be $\sim(2.09\pm0.17)\times 10^{42} $ erg $
\rm s^{-1}$ with the distance of 74.7 Mpc from \citet{Gao2004}, which was approximately consistent with that of \citet{Peng2016}.

Based on the above analysis, we suggest that the variability of the GeV $\gamma$-ray emission of the northwest region of Arp 220 was  probably from CRATES J153246+234400, and Arp 220 was a stable source in the current observation period.

\section{discussion and conclusion}

In this study, we reexplored the origin of the unknown $\gamma$-ray excess from the northwest region of Arp 220 by analyzing $\sim$11.8 years of \textit{Fermi}-LAT data. 
By analyzing its spatial position, we found that CRATES J153246+234400 lay in the 1$\sigma$ error circle of the best-fit position of the unknown $\gamma$-ray excess.
Comparing our previous results with those of \citet{Peng2016}, we found that the photon flux in the 0.2-100 GeV band was approximately two times larger, and the significance level was approximately four times higher than previous research results, which suggests the likely variability from the GeV $\gamma$-ray emission of CRATES J153246+234400. 

Then, the $\gamma$-ray emission from the $\gamma$-ray excess was reanalyzed in our work. 
First, the LCs with 15 and 45 time bins for the whole time showed two active periods.
Furthermore, we that found a constant flux model could not fit these LCs well, particularly for the LC of 45 time bins with a large reduced $\chi^{2} $ value. 
Finally, by calculating $\rm TS_{var}$ for the two LCs of 15 and 45 time bins, we found that the LC of 45 time bins had a  large $\rm TS_{var}$ value corresponding to the significance level of 5.94$\sigma$, which further indicated that the variability of the $\gamma$-ray emission existed at the  location of CRATES J153246+234400. 
Comparing the two different active periods of P3 and P4 from the LC of 45 time bins, we found that the variability of P4 was more significant than that of P3, and the position of CRATES J153246+234400 was always within the 1$\sigma$ error circle of the two different periods of P3 and P4, which further strongly indicated that the $\gamma$-ray radiations of the two active periods were likely to come from CRATES J153246+234400, excluding AGNs closer to the region of the $\gamma$-ray excess (e.g., SDSS J15323+2345, SDSS J15337+2358, and  1WGA J1533.8+2356).
  
 Additionally, we found that the ranges of the spectral  indices for the whole time for P3 and P4 were all within the average index range of 2.4-2.5 of $\gamma$-ray FSRQs \citep[e.g.,][]{Ackermann2015}, which suggested that the spectral characteristics of CRATES J153246+234400 were consistent with the previously observed FSRQs previously observed.
 Considering Arp 220, no significant variability in the $\gamma$-ray emission was found in this analysis.

\section{Acknowledgements}
We sincerely appreciate the support for this work from the National Key R\&D Program of China under Grant No.2018YFA0404204, the National Natural Science Foundation of China (NSFC U1931113, U1738211) ,the Foundations of Yunnan Province (2018IC059, 2018FY001(-003)), the Scientific research fund of Yunnan Education Department
(2020Y0039).


\end{document}